\documentclass[aps,pre,english,showpacs,showkeys,twocolumn,preprintnumbers,floatfix,nofootinbib,10pt]{revtex4-2}
\usepackage{eurosym}
\usepackage{amssymb}
\usepackage{graphicx,color}
\usepackage{epsfig}
\usepackage{rotating}
\usepackage[T1]{fontenc}
\usepackage{latexsym,epsfig}
\usepackage[latin9]{inputenc}
\usepackage{amsfonts}
\usepackage{bm}
\usepackage{babel}
\usepackage{hyperref}
\usepackage{amsmath,mathrsfs}
\usepackage{xcolor}

\pdfoutput=1
\begin{document}

\title{Revisiting the Contact Model with Diffusion Beyond the Conventional
Methods}
\author{Roberto da Silva\textsuperscript{1},  E. Venites Filho\textsuperscript{1}, Henrique A. Fernandes\textsuperscript{2}, Paulo F. Gomes\textsuperscript{2}}
\address{1 - Instituto de F{\'i}sica, Universidade Federal do Rio Grande do Sul,
Av. Bento Gon{\c{c}}alves, 9500 - CEP 91501-970, Porto Alegre, Rio Grande do
Sul, Brazil\\ 
2 - Instituto de Ci{\^e}ncias Exatas e Tecnol{\'o}gicas, Universidade Federal de Jata{\'i}, BR 364, km 192, 3800 - CEP 75801-615, Jata{\'i}, Goi{\'a}s, Brazil}

\begin{abstract}
The contact process is a non-equilibrium Hamiltonian model that, even in one
dimension, lacks an exact solution and has been extensively studied via
Monte Carlo simulations, both in steady-state and time-dependent scenarios.
Although the effects of particle mobility/diffusion on criticality have been
preliminarily investigated, they remain incompletely understood. In this
work, we examine how the critical rate of the model varies with the
probability of particle mobility. By analyzing different stochastic
evolutions of the system, we employ two modern approaches: 1) Random Matrix
Theory (RMT): By building on the success of RMT, particularly Wishart-like
matrices, in studying statistical physics of systems with up-down symmetry
via magnetization dynamics [R. da Silva, IJMPC 2022], we demonstrate its
applicability to models with an absorbing state. 2) Optimized Temporal Power
Laws: By using short-time dynamics, we optimize power laws derived from
ensemble-averaged evolutions of the system. Both methods consistently reveal
that the critical rate decays with mobility according to a simple Belehradek
function. Additionally, a straightforward mean-field analysis supports the
decay of the critical parameter with mobility, although it predicts a
simpler linear dependence.
\end{abstract}

\maketitle

\section{Introduction}

\label{Sec:Introduction}

Models with absorbing states exhibit scaling behavior at criticality, marking the transition between active and absorbing stationary states despite the absence of a Boltzmann distribution governing the evolution of the systems. A fundamental example is the standard contact process, which undergoes a phase transition even in one-dimensional systems. Proposed by Harris \cite
{Harris} and extensively analyzed by Liggett \cite{Ligget}, this model does not has an exact analytical solution on a lattice, including the simplest one-dimensional case. So, it is not for nothing, its critical properties continue to hold significant relevance in statistical physics \cite{Dickman,Hinrichsen2000,Henkel}.

The contact process (CP) can be interpreted on the lattice in two distinct ways: as empty (inactive) sites or sites occupied by particles (active sites) that can undergo creation/annihilation processes, or as a specific implementation of the susceptible-infected-susceptible (SIS) epidemic model. In that case, the infection dynamics follows these simple rules:

1. When a healthy (susceptible) site has two infected nearest neighbors, the probability of infection is: $p_{(\bullet \circ \bullet )\rightarrow (\bullet \bullet \bullet )}=\lambda (1+\lambda )^{-1}$.

2. If only one nearest neighbor is infected, the probability is halved: $p_{(\circ \circ \bullet )\rightarrow (\circ \bullet \bullet )}=p_{(\bullet \circ \circ )\rightarrow (\bullet \bullet \circ )}=\lambda (1+\lambda )^{-1}/2$.

3. Regardless of the neighboring states, an infected site recovers with probability: $p_{(\bullet )\rightarrow (\circ )}=(1+\lambda )^{-1}$. 

When considering the creation/annihilation process, the steady-state density of active sites, $\overline{\rho }$, exhibits critical behavior:

a) \textbf{Absorbing phase}: For $\lambda < \lambda _{c}$, $\overline{\rho } = 0$, indicating that the system falls into an absorbing state where the entire population remains healthy and the dynamics cease.

b) \textbf{Active phase}: For $\lambda > \lambda_{c}$, the system reaches an active phase characterized by $\overline{\rho }= \left( \lambda -\lambda _{c}\right) ^{\beta }$.

Several important studies have investigated this model, including: determination of the critical parameter $\lambda_c$ through series expansion analysis \cite{Dickman1989}, examination of universality in both local and global two-time density correlators and their associated linear response functions \cite{Bottcher}, application of mixed initial conditions to extract critical exponents \cite{Silva2004} among other significant contributions to the field.

However, from our point of view, there is still an important aspect that warrants further scientific investigation in the contact process: the role of particle diffusion. We emphasize that the CP inherently contains diffusion mechanisms. Consider,
for example, the transitions: 
\begin{align*}
11101 &\rightarrow 11001 \quad \text{(annihilation at site 3)} \\
11001 &\rightarrow 11011 \quad \text{(creation at site 4)}
\end{align*}
These consecutive transitions effectively represent the diffusion of a particle from position 3 to position 4, despite being mediated through annihilation-creation dynamics.

Our question, in contrast, focuses on the effects of standard diffusion rather than those resulting from rules of creation and annihilation. It is unsurprising that numerous intriguing studies have explored diffusion in the one-dimensional contact process within the literature. Dantas, Oliveira, and Stilck \cite{Dantas} employed supercritical series expansion and finite-size
exact solution approaches to examine the impact of diffusion on criticality, determining an existence of a multicritical point. They also estimated the dependence of the crossover exponent on the diffusion rate \cite{Dantas2}. Shen et al. \cite{Shen}, on the other hand, used machine learning techniques to investigate the pair-contact process with diffusion.

In this study, we investigate how the critical exponent $\lambda_c$ depends on the diffusion probability $\gamma$, the probability which a given site has to swape the states of neighboring sites. In order to carry out our study, we employ two distinct and highly effective methods, both utilizing time-dependent Monte Carlo (MC) simulations. The first method refines the power-law approach by applying the coefficient of determination to the linearization of the power laws, as originally introduced in Ref. \cite{Silva2012}. The second method involves the analysis of the spectra of correlation random matrices derived from short-time series, as detailed in Ref. \cite{RMT2023}.

It is worth noting that in this second method, our focus extends further. In particular, we aim to investigate the behavior of the spectra of certain random matrices constructed through the time evolution of absorbing state models, such as the contact process. This analysis is novel for this class of nonequilibrium models, since previous studies have only addressed systems with up-down symmetry, such as the Kinetic Ising model \cite{RMT2023}.

Before presenting our results in the next section, we revisit the mean-field modeling of the contact process (CP) through an empirical study, which provides insights into the expectations for our one-dimensional simulations. In Section \ref{Sec:Methods}, we outline the methods used to determine the critical rate $\lambda_c$ as a function of the diffusion/mobility probability $\gamma$. Finally, in Section \ref{Sec:Results}, we present our findings, followed by a brief summary and conclusions in Section \ref{Sec:Conclusions}.

\section{Mean-field approximation and pedagogical arguments}

\label{Sec:Mean-Field}

In one dimension, the transition rate can be expressed as \cite{TaniaMario2014}: 
\begin{equation}
\begin{array}{lll}
\omega (\sigma _{i}\rightarrow 1-\sigma _{i}|\sigma _{i-1},\sigma _{i+1}) & =
& \frac{\lambda }{2}(1-\sigma _{i})(\sigma _{i-1}+\sigma _{i+1}) \\ 
&  & +\ \sigma _{i}%
\end{array}
\label{Eq:Transition}
\end{equation}%
for the site $i $ from state $\sigma_i $ to state $1 - \sigma_i $, where $%
\sigma_i = 1 $ ($\bullet $) denotes an active site and $\sigma_i = 0 $ ($%
\circ $) denotes an inactive site. We can verify that $\omega (\circ
\rightarrow \bullet | \bullet, \bullet ) = \lambda $ and $\omega (\circ
\rightarrow \bullet | \circ, \bullet ) = \omega (\circ \rightarrow \bullet |
\bullet, \circ ) = \lambda / 2 $ and $\omega (\bullet \rightarrow \circ ) =
1 $. When these rates are converted into probabilities, we obtain exactly
the probabilities of the CP, since $p_{(\bullet \circ \bullet ) \rightarrow
(\bullet \bullet \bullet )} = \frac{\omega (\circ \rightarrow \bullet |
\bullet, \bullet )}{Z} $, $p_{(\circ \circ \bullet ) \rightarrow (\circ
\bullet \bullet )} = \frac{\omega (\circ \rightarrow \bullet | \circ,
\bullet )}{Z} = p_{(\bullet \circ \circ ) \rightarrow (\bullet \bullet \circ
)} = \frac{\omega (\circ \rightarrow \bullet | \bullet, \circ )}{Z} $, and $%
p_{(\bullet ) \rightarrow (\circ )} = \frac{\omega (\bullet \rightarrow
\circ )}{Z} $, which gives exactly the probabilities reported in the first
section.

The master equation says that:%
\begin{equation*}
\begin{array}{ccc}
\frac{dP(\sigma ,t)}{dt} & = & \sum\limits_{i=1}^{L}\left[ \omega \left(
1-\sigma _{i}\rightarrow \sigma _{i}\right) \Pr \left( \sigma ^{\prime
},t\right) \right. \\ 
&  & \left. -\omega \left( \sigma _{i}\rightarrow 1-\sigma _{i}\right) \Pr
\left( \sigma ,t\right) \right]%
\end{array}%
\end{equation*}%
with $\sigma \equiv (\sigma _{1},\sigma _{2},...\sigma _{i-1},\sigma
_{i},\sigma _{i+1},...,\sigma _{N})$ and $\sigma ^{\prime }\equiv \left(
\sigma _{1},\sigma _{2},...\sigma _{i-1},(1-\sigma _{i}),\sigma
_{i+1},...,\sigma _{N}\right) $, where similarly $\omega \left( 1-\sigma
_{i}\rightarrow \sigma _{i}\right) =\frac{\lambda }{2}\sigma _{i}(\sigma
_{i-1}+\sigma _{i+1})+\left( 1-\sigma _{i}\right) $. By saving notation, we
removed the conditional symbol of Eq. (\ref{Eq:Transition}). We have $%
\left\langle \sigma _{j}\right\rangle =\sum\limits_{\sigma }\sigma
_{j}P(\sigma ,t)$ and differentiating it with respect to time, we obtain: 
\begin{equation*}
\begin{array}{lll}
\frac{d\left\langle \sigma _{j}\right\rangle }{dt} & = & \sum\limits_{\sigma
}\sigma _{j}\frac{dP(\sigma ,t)}{dt} \\ 
&  &  \\ 
& = & \sum\limits_{i=1}^{L}\sum\limits_{\sigma }\sigma _{j}\omega \left(
1-\sigma _{i}\rightarrow \sigma _{i}\right) \Pr \left( \sigma ^{\prime
},t\right) \\ 
& - & \sum\limits_{i=1}^{L}\sum\limits_{\sigma }\sigma _{j}\omega \left(
\sigma _{i}\rightarrow 1-\sigma _{i}\right) \Pr \left( \sigma ,t\right) \\ 
&  &  \\ 
&  &  \\ 
& = & \sum\limits_{\sigma }(1-\sigma _{j})\omega \left( \sigma
_{j}\rightarrow 1-\sigma _{j}\right) \Pr \left( \sigma ,t\right) \\ 
& - & \sum\limits_{\sigma }\sigma _{j}\omega \left( \sigma _{j}\rightarrow
1-\sigma _{j}\right) \Pr \left( \sigma ,t\right) \\ 
&  &  \\ 
& = & \left\langle (1-2\sigma _{j})\omega \left( \sigma _{j}\rightarrow
1-\sigma _{j}\right) \right\rangle.
\end{array}%
\end{equation*}

Thus,
\begin{equation}
\begin{array}{lll}
\frac{d\left\langle \sigma _{j}\right\rangle }{dt} & = & \left\langle
(1-2\sigma _{j})\left( \frac{\lambda }{2}(1-\sigma _{j})(\sigma
_{j-1}+\sigma _{j+1})+\sigma _{j}\right) \right\rangle \\ 
&  &  \\ 
& = & \left\langle \sigma _{j}-2\sigma _{j}^{2}+\frac{1}{2}\lambda \sigma
_{j-1}+\frac{1}{2}\lambda \sigma _{j+1}-\frac{3}{2}\lambda \sigma _{j}\sigma
_{j-1}\right. \\ 
&  &  \\ 
&  & \left. -\allowbreak \frac{3}{2}\lambda \sigma _{j}\sigma _{j+1}+\lambda
\sigma _{j}^{2}\sigma _{j-1}+\lambda \sigma _{j}^{2}\sigma
_{j+1}\right\rangle.
\end{array}%
\end{equation}

For the CP model, we have: 
\begin{equation}
\left\langle \sigma _{j}^{2}\right\rangle =\left\langle \sigma
_{j}\right\rangle =\rho,
\end{equation}
and, in addition, we have
\begin{equation}
\left\langle \sigma _{j}^{2}\sigma _{j-1}\right\rangle =\left\langle \sigma
_{j}\sigma _{j-1}\right\rangle \text{ or }\left\langle \sigma _{j}^{2}\sigma
_{j+1}\right\rangle =\left\langle \sigma _{j}\sigma _{j+1}\right\rangle
\end{equation}.

By a simple approximation, one leads to: 
\begin{equation}
\left\langle \sigma _{j}\sigma _{j-1}\right\rangle \approx \left\langle
\sigma _{j}\right\rangle \left\langle \sigma _{j-1}\right\rangle =\rho
^{2}\approx \left\langle \sigma _{j}\sigma _{j+1}\right\rangle.
\end{equation}

Thus, we obtain
\begin{equation}
\frac{d\rho }{dt}=(\lambda -1)\rho -\lambda \rho^{2}.
\label{Eq:previous_CP_MF}
\end{equation}

Finally, we incorporate the external diffusive effects into the equation. To this end, we adopt an empirical approach to illustrate how such effects might manifest. At low densities, diffusion should be highly sensitive to increases in the density of active sites, as greater mobility enhances the chances for inactive (or healthy) particles to become active (or infected). This effect is expected to grow proportionally with density up to a certain point, beyond which mobility becomes less sensitive, approaching saturation as $\rho \to 1$. Naturally, no diffusion is expected at the extremes $\rho = 0$ and $\rho = 1$. These considerations motivate the introduction of an empirical term of the form:
\begin{equation}
h_{\text{diff}}(\rho) = D \rho (1 - \rho)  \label{Eq:external_diffusion}
\end{equation}
in Eq. (\ref{Eq:previous_CP_MF}), where $D$ is a constant related to the mobility of the particles, that is, the probability of diffusion.

In a recent study, some authors revealed that in nightclub dynamics \cite{nightclub}, 
there is an optimal crowd density to maximize profits, as people can reach the bar counter more easily due to faster diffusion.
This is directly related to a term of this kind, as it reflects the
efficiency of customer mobility within the nightclub, thus motivating our approach. So, we add this term \textit{ad hoc} in Eq. (\ref{Eq:previous_CP_MF}) with the following result:

\begin{equation}
\begin{array}{lll}
\frac{d\rho }{dt} & = & (\lambda -1)\rho -\lambda \rho ^{2}+D\rho (1-\rho )
\\ 
&  &  \\ 
& = & \left( \lambda -1+D\right) \rho -(D+\lambda )\rho ^{2}%
\end{array}%
\end{equation}

The solution of this equation is given by:%
\begin{equation}
\rho (t)=\frac{\left( \lambda -1+D\right) \rho _{0}e^{\left( \lambda
-1+D\right) t}}{\left[ \lambda -1+D-(D+\lambda )\rho _{0}\right] +(D+\lambda
)\rho _{0}e^{\left( \lambda -1+D\right) t}}  \label{Eq:Exponential}
\end{equation}%
where $\rho _{0}=\rho (0)$. In the steady state, $\frac{d\rho }{dt}=0$ and
then: \ 
\begin{equation*}
\rho _{\infty }=\frac{\left( \lambda -1+D\right) }{(D+\lambda )}
\end{equation*}

Thus we can estabilish a dependence of $\lambda _{C}$ on $D:$%
\begin{equation}
\lambda _{C}=1-D  \label{Eq:Mean-field-main-equation}
\end{equation}%
which suggests that when $D$ increases $\lambda _{C}$ decreases. It is
interesting to observe that in the vicinity of the criticality: $\lambda
\approx \lambda _{C}=1-D$, and therefore: $e^{\left( \lambda -1+D\right)
t}\approx 1+\left( \lambda -1+D\right) t:$%
\begin{equation*}
\begin{array}{lll}
\rho (t) & \approx & \frac{\left( \lambda -1+D\right) \rho _{0}\left[
1+\left( \lambda -1+D\right) t\right] }{\lambda -1+D+(D+\lambda )\rho
_{0}\left( \lambda -1+D\right) t} \\ 
&  &  \\ 
& \approx & \frac{\rho _{0}}{1+\frac{(D+\lambda )}{\left( \lambda
-1+D\right) }\rho _{0}\left( \lambda -1+D\right) t} \\ 
&  &  \\ 
& \sim & t^{-1}\text{, when }t\rightarrow \infty%
\end{array}%
\end{equation*}

This simple mean-field analysis aims to determine how $\lambda_{C}$ depends on the diffusion in the case of short-range, one-dimensional contact. In this work, we study the CP model by introducing diffusion into the nearest-neighbor CP defined in the previous section. Our diffusion model randomly selects a site, and a replacement with its nearest neighbor is performed with a probability $0 \leq \gamma \leq 1$, where $\gamma$ is related to $D$ in the mean-field regime. The goal is to explore how $\lambda_{C}$ decays as a function of $\gamma$. However, unlike the simple linear relationship in Eq. (\ref{Eq:Mean-field-main-equation}), this paper will address the nature of this decay.

To determine such behavior, we will use two techniques:

\begin{enumerate}
\item Optimization of the coefficient of determination of temporal power laws, proposed in Ref. \cite{Silva2012};

\item Spectral analysis of correlation matrices constructed from multiple realizations of the contact process, following the method presented in Ref. \cite{RMT2023}.
\end{enumerate}

We investigated whether both methods, which rely on the time evolution of the active site density, produce the same underlying physics whereas the first approach considers averages over different time series, while the second one examines the spectral properties of matrices built from magnetization trajectories.

\section{Simulation and Methods}

\label{Sec:Methods}

In this work, our order parameter is the density of active particles $\rho$ which, for the $i$-th sample and at the $t-$th Monte Carlo (MC) step, can be defined by the matrix element: 
\begin{equation}
\rho _{i,t}=\frac{1}{L}\sum_{k=1}^{L}s_{k}^{(i)}(t),  \label{eq:density}
\end{equation}%
where $s_{k}^{(i)}(t)$ is 0 if the site $k$ is inactive and 1 otherwise, and $L$ is the size of the system. From this quantity, we explore different approaches to study the one-dimensional contact process with diffusion by means of nonequilibrium MC simulations. At each step, a site is chosen randomly with probability $
1/L$. If the site is active, it may become inactive, and if inactive, it may become active depending on its neighborhood, following the transition probabilities defined in Section \ref{Sec:Introduction}. One MC step consists of $L$ randomly chosen sites undergoing possible transitions. Next, we account for particle diffusion: In this case, $L$ sites are also randomly selected, each one with the
same probability $1/L$. To address the diffusion process, each chosen site exchanges position with one of its two nearest neighbors (chosen at random) with probability $\gamma$.

We perform a total of $N_{sample}$ independent time evolutions, each
consisting of $N_{steps}$ MC steps.

To determine $\lambda _{c}$ as a function of $\gamma $, we employ two
methods. The first one is based on optimizing power laws \cite{Silva2012}, where the best fit of the power-law occurs at $\lambda =\lambda _{c}$. The second one is aspectral method, which identifies $\lambda _{c}$ as the value where the average eigenvalue of correlation matrices reaches a minimum, following the methodology described in Ref. \cite{RMT2023}.

\subsection{First method: Optimization of power laws by the coefficient of determination}

Our first method involves averaging over different system evolutions or
across $N_{samples}$ independent simulations:

\begin{equation}
\left\langle \rho (t)\right\rangle =\frac{1}{N_{sample}}\sum%
\nolimits_{i=1}^{N_{sample}}\rho _{i,t}
\end{equation}

Unlike equilibrium systems, where spatial scaling is often the primary
concern, nonequilibrium systems require both temporal and spatial scaling to
be considered. This is particularly evident in systems with absorbing
states, where a universal scaling relation emerges. This relation, which has
significant implications, dictates that the density of active sites must
follow the scaling law dependent on initial conditions \cite%
{Dickman,Hinrichsen2000,Henkel}:

\begin{equation}
\left\langle \rho (t)\right\rangle \sim t^{-\beta /\nu _{\parallel
}}f((\lambda -\lambda _{c})t^{1/\nu _{\parallel }},t^{d/z}L^{-d},\rho
_{0}t^{\beta /\nu _{\parallel }+\theta }),  \label{Eq:scaling_temporal}
\end{equation}%
where $\rho _{0}=\rho (0)$ is the initial density of active sites, and $d$
is the spatial dimension. The indices $z=\nu _{\parallel }/\nu _{\perp }$
and $\theta =\frac{d}{z}-\frac{2\beta }{\nu _{\parallel }}$ are dynamic
critical exponents, while $\beta $, $\nu _{\parallel }$, and $\nu _{\perp }$
are static critical ones.

The quantity $\lambda -\lambda _{c}$ represents the distance from the
critical point and governs the algebraic behavior of two independent
correlation lengths: the spatial correlation length, $\xi _{\perp }\sim
(\lambda -\lambda _{c})^{-\nu _{\perp }}$ and the temporal correlation
length, $\xi _{\parallel }\sim (\lambda -\lambda _{c})^{-\nu _{\parallel }}$%
. At criticality ($\lambda =\lambda _{c}$), different critical exponents can
be estimated from Eq. (\ref{Eq:scaling_temporal}) by considering
different initial conditions at the beginning of the simulations ($t=0$). In the context of time-dependent MC simulations, starting from an initial density $\rho =\rho _{0}$, the following crossover of power laws is expected:

\begin{equation}
\left\langle \rho (t)\right\rangle \sim \left\{ 
\begin{array}{ll}
\rho _{0}t^{\theta } & \text{if }t_{mic}<t<\rho _{0}^{-\frac{\nu _{\parallel
}}{d\ \nu _{\perp }-\beta }} \\ 
&  \\ 
t^{-\delta } & \text{if }\rho _{0}^{-\frac{\nu _{\parallel }}{d\ \nu _{\perp
}-\beta }}\leq t\text{,}%
\end{array}%
\right.  \label{Eq:Crossover}
\end{equation}%
with 
\begin{equation}
\text{ }\delta =\frac{\beta }{\nu _{\parallel }}\text{.}
\label{eq:exponents}
\end{equation}

We focus on the second power law in Eq. (\ref{Eq:Crossover}). To derive this
power law directly, we begin by initializing the system with $\rho _{0}=1$.
In this case, the decay follows a power law of the form $\left\langle \rho
(t)\right\rangle \sim t^{-\delta }$, which is observed at $\lambda =\lambda
_{c}$. For $\lambda \neq \lambda _{c}$, an exponential behavior is observed,
consistent with the mean-field analysis which we developed above and summarized in Eq. (\ref{Eq:Exponential}).

To pinpoint the critical parameters, we employ a straightforward yet highly effective statistical method for monitoring temporal power laws within time-dependent MC simulations: the coefficient of determination (COD) \cite{trivedi2002}. This approach recognizes that there are two sources of variation when fitting experimental data linearly: the
explained variation and the unexplained one. The total variation is
the sum of both variations, and the quality of the linear fit is quantified by the ratio of explained variation to total variation, yielding a COD value (denoted as $r$) between 0 and 1. A higher COD value indicates a better linear fit.

The method used in this work is based on the approach introduced by da
Silva, Drugowich, and Martinez in their study of nonequilibrium statistical mechanics \cite{Silva2012}. In that work, they employed the COD in the linearization of the time evolution of averaged magnetizations to localize the critical parameters of the generalized Ising model. Although originally developed for this specific context, the method is generalizable and has been applied successfully to various other models (see, for example, \cite{silva2014}).

We start by computing the average log-densities, $\left\langle \ln \rho
(t|\lambda ,\gamma )\right\rangle $, across multiple runs. For each
parameter pair $(\lambda ,\gamma )$, we then determine the coefficient of
determination, $r$, as follows: 
\begin{equation}
r(\lambda ,\gamma )=\frac{\sum\limits_{t=N_{\min }}^{N_{steps}}(\overline{%
\ln \left\langle \rho (t|\lambda ,\gamma )\right\rangle }-a-b\ln t)^{2}}{%
\sum\limits_{t=N_{\min }}^{N_{steps}}(\overline{\ln \left\langle \rho
(t|\lambda ,\gamma )\right\rangle }-\ln \left\langle \rho \right\rangle
(t))^{2}}.  \label{eq:coef_det}
\end{equation}%
Here, the overline indicates an additional averaging step: 
\begin{equation}
\overline{\ln \left\langle \rho (t|\lambda ,\gamma )\right\rangle }=\frac{1}{%
N_{steps}}\sum\nolimits_{t=N_{\min }}^{N_{steps}}\ln \left\langle \rho
\right\rangle (t).
\end{equation}

The values $N_{steps}$ and $N_{\min }$ correspond to the total number of
Monte Carlo (MC) steps performed and the number of initial steps discarded
during equilibration, respectively. The optimal parameters are then
determined via marginalization: 
\begin{equation}
\lambda _{c}(\gamma )\equiv \arg \max_{\lambda _{\min }\leq \lambda \leq
\lambda _{\max }}\{r(\lambda |\gamma )\},
\end{equation}
that is, for a fixed $\gamma $, we vary $\lambda $ to find the value that
maximizes $r$. The parameters $a$ and $b$ correspond to the intercept and
slope, respectively, of the linear function. When initialized with $\rho
_{0}=1$, $b$ equals $\delta $ at the critical point. The intervals $[\lambda
_{\min },\lambda _{\max }]$ are selected to ensure proper refinement.

As defined in Eq. (\ref{eq:coef_det}), $r$ quantifies the ratio of explained
variation to total variation. Near the critical point $\lambda =\lambda _{c}$%
, $r$ approaches unity ($r\approx 1$), indicating that $\rho (t)$ follows a
power-law distribution. This power-law behavior manifests as a linear
relationship on a log-log scale. Conversely, far from the critical point, $%
\rho (t)$ deviates from power-law scaling, as described in Eq. (\ref%
{Eq:Crossover}), resulting in $r\approx 0$.

\subsection{Second Method: Spectral Analysis of Correlation Matrices}

The second method builds on the methodology introduced in Ref. \cite{RMT2023}, where a cross-correlation matrix $\mathcal{C}$ is constructed from time series iterations of a given observable---such as generalized momentum, generalized position, or magnetization per particle. Although prior works, including Refs. \cite{RMT2023,Vinayak2014,Biswas2017}, have explored the method by using the eigenvalues of such matrices to detect phase transitions in spin systems, this technique has also proven to be effective in quantifying genuine correlations in financial markets. In Econophysics, for instance, financial returns serve as the observable \cite{Stanley,Stanley2,Stanleyb,Bouchaud,Bouchaud2}.

Interestingly, the foundational approach traces back to Wishart's 1928 work \cite{Wishart} on what later became known as Wishart ensembles in
Statistics---predating the famous random matrix ensembles of Wigner and
Dyson \cite{Wigner,Dyson,Mehta} in Statistical Mechanics. Despite its long history, this methodology still holds unexplored potential for novel applications.

In this work, we instead focus on the density of active sites as the
observable. The elements of the cross-correlation matrix $\mathcal{C}$ are defined as:

\begin{equation}
\mathcal{C}_{ij}\mathcal{=}\frac{\left\langle \rho _{i}\rho
_{j}\right\rangle -\left\langle \rho _{i}\right\rangle \left\langle \rho
_{j}\right\rangle }{\sigma _{\rho _{i}}\sigma _{\rho _{j}}}
\end{equation}%
where the averages and standard deviations are calculated from two time series of length $N_{steps}$: $\rho _{k,0},\rho _{k,1},...,\rho _{k,N_{steps-1}}$, where $k=i,j=1,...,N_{sample}$.

The expectation values are given by:

\begin{eqnarray}
\left\langle \rho _{k}\right\rangle &=&\frac{1}{N_{steps}}%
\sum_{p=0}^{N_{steps-1}}\rho _{kp}\text{, } \\
&&  \notag \\
\left\langle \rho _{i}\rho _{j}\right\rangle &=&\frac{1}{N_{steps}}%
\sum_{p=0}^{N_{steps}-1}\rho _{ip}\rho _{jp}\text{ }  \notag
\end{eqnarray}%
and the standard deviation $\sigma _{\rho _{k}}$ is calculated via 
\begin{equation*}
\sigma _{\rho _{k}}=\frac{1}{(N_{steps}-1)}\sum_{p=0}^{N_{steps}-1}(\rho
_{kp}-\left\langle \rho _{k}\right\rangle )^{2}\approx \left\langle \rho
_{k}^{2}\right\rangle -\left\langle \rho _{k}\right\rangle ^{2}\text{.}
\end{equation*}

It is important to note that the matrix $\mathcal{C}$, of dimension $%
N_{sample}$, can be obtained from the standardized time-evolution matrix\ $%
\mathcal{M}$:

\begin{widetext}
\begin{equation*}
\mathcal{M}=\left( 
\begin{array}{cccc}
\frac{\rho _{10}-\left\langle \rho _{1}\right\rangle }{\sigma _{\rho _{1}}}
& \frac{\rho _{20}-\left\langle \rho _{2}\right\rangle }{\sigma _{\rho _{2}}}
& \cdots & \frac{\rho _{N_{sample}0}-\left\langle \rho
_{N_{sample}}\right\rangle }{\sigma _{\rho _{N_{sample}}}} \\ 
\frac{\rho _{11}-\left\langle \rho _{1}\right\rangle }{\sigma _{\rho _{1}}}
& \frac{\rho _{21}-\left\langle \rho _{2}\right\rangle }{\sigma _{\rho _{2}}}
&  & \frac{\rho _{N_{sample}1}-\left\langle \rho _{N_{sample}}\right\rangle 
}{\sigma _{\rho _{N_{sample}}}} \\ 
\vdots & \vdots &  & \vdots \\ 
\frac{\rho _{1N_{steps}-1}-\left\langle \rho _{1}\right\rangle }{\sigma
_{\rho _{1}}} & \frac{\rho _{2N_{steps}-1}-\left\langle \rho
_{2}\right\rangle }{\sigma _{\rho _{2}}} &  & \frac{\rho
_{N_{sample}N_{steps}-1}-\left\langle \rho _{N_{sample}}\right\rangle }{%
\sigma _{\rho _{N_{sample}}}}%
\end{array}%
\right)
\end{equation*}
\end{widetext}

This allows us to verify that:

\begin{equation}
\mathcal{C=}\frac{1}{N_{steps}}\mathcal{M}^{t}\mathcal{M}.
\label{Eq:Covariance_matrix}
\end{equation}

It is known that if $\rho _{i,0},\rho _{i,1},...,\rho _{i,N_{steps-1}}$ are
a set of independent random variables, we are in the context of the real
Wishart ensemble \cite{Seligman3,Novaes}. In that case, the joint
probability distribution of the eigenvalues is expected to be given by:

\begin{equation}
P(E_{1},...,E_{N_{sample}})=\frac{1}{Z}e^{-\mathcal{H}%
(E_{1},...,E_{N_{sample}})}\text{,}
\end{equation}%
where 
\begin{equation}
Z=\idotsint dE_{1}...dE_{N_{sample}}e^{-\mathcal{H}%
(E_{1},...,E_{N_{sample}})}
\end{equation}%
and

\begin{equation}
\begin{array}{lll}
\mathcal{H}(E_{1},...,E_{N_{sample}}) & = & \mathcal{H}_{\text{auto}%
}(E_{1},...,E_{N_{sample}})+ \\ 
&  &  \\ 
&  & \mathcal{H}_{\text{int}}(E_{1},...,E_{N_{sample}}),
\end{array}%
\end{equation}%
corresponding to a Hamiltonian of a Coulomb gas with logarithmic repulsion 
\begin{equation}
\mathcal{H}_{\text{int}}(E_{1},...,E_{N_{sample}})=-\sum_{i<j}\ln \left\vert
E_{i}-E_{j}\right\vert \text{,}  \label{Eq:int}
\end{equation}%
and the term 
\begin{equation}
\begin{array}{ll}
\mathcal{H}_{\text{auto}}(E_{1},...,E_{N_{sample}}) & =\frac{N_{steps}}{2}%
\sum_{i=1}^{N_{sample}}E_{i} \\ 
&  \\ 
& -\frac{(N_{steps}-N_{sample}-1)}{2}\sum_{i=1}^{N_{sample}}\ln E_{i}\text{,}%
\end{array}
\label{Eq:auto}
\end{equation}%
attracts the particles to the origin, with $E_{1},...,E_{N_{sample}}\geq 0$.
Since $N_{steps}>N_{sample}$, we have 
\begin{equation*}
E>\frac{(N_{steps}-N_{sample}-1)}{N_{steps}}\ln E\text{.}
\end{equation*}

In the case of potentials described by Eqs. (\ref{Eq:int}) and (\ref{Eq:auto}), the density of states is defined by: 
\begin{equation}
\sigma (E)=\int_{0}^{\infty }\int_{0}^{\infty }...\int_{0}^{\infty
}P(E,E_{2}...,E_{N_{sample}})\prod_{i=1}^{N_{sample}}dE_{i}
\end{equation}%
following the well-known Marchenko-Pastur (MP) law \cite{Marcenko,Sengupta}:

\begin{equation}
\sigma _{M}(E)=\left\{ 
\begin{array}{l}
\dfrac{N_{MC}}{2\pi N_{sample}}\dfrac{\sqrt{(E-E_{-})(E_{+}-E)}}{E}\ \text{%
if\ }
\\
\ \ \ \ \ \ \ \ \ \ \ \ \ \ \ \ \ \ \ \ \ \ E_{-}\leq E\leq E_{+} \\ 
\\ 
0\ \ \ \ \text{otherwise,}%
\end{array}%
\right.  \label{Eq:MP}
\end{equation}%
where 
\begin{equation*}
E_{\pm }=1+\frac{N_{sample}}{N_{MC}}\pm 2\sqrt{\frac{N_{sample}}{N_{MC}}}%
\text{.}
\end{equation*}

We construct matrices by examining the evolution of active sites of the CP model. A key quantity to track is the deviation of the density of states from the non-correlated scenario, as defined by $\sigma _{M}(E)$ in Eq. (\ref{Eq:MP}). This deviation is captured by the first moment of $\sigma (E)$, given by $E[E]=\int_{0}^{\infty }E\ \sigma (\lambda )d\lambda $. Its estimator is derived from the numerical density of states:

\begin{equation}
\left\langle E\right\rangle =\frac{\sum_{i=1}^{N_{int}}\sigma
_{N}(E_{i})E_{i}}{\sum_{i=1}^{N_{int}}\sigma _{N}(E_{i})}\text{,}
\label{Eq:Average}
\end{equation}%
where $N_{int}$ represents the number of bins used to compute the numerical density of states $\sigma _{N}(E_{i})$, which may differ from the expression in Eq. \ref{Eq:MP}. This leads to a change in the function $\mathcal{H}_{\text{auto}}(E_{1},...,E_{N_{sample}})$ from that given in Eq. (\ref{Eq:MP}), introducing gaps in the density of states. These gaps are reflected in the first moment $\left\langle E\right\rangle $ or higher moments $\sigma (E)$. Our previous work has demonstrated that extreme values of $\left\langle E\right\rangle $ are closely associated with phase transition points in spin systems \cite{RMT2023} and chaotic-to-stability transitions
in chaotic maps \cite{RMT2023-4}.

An important distinction here is that, unlike Ising systems, where we
expect $\sigma _{N}(E)\rightarrow \sigma _{M}(E)$ as $T\rightarrow \infty $, as described in Eq. (\ref{Eq:MP}), or at least a good approximation at this limit, as shown in Ref. \cite{RMT2023}, this does not necessarily hold for the CP model. In the CP model, as $\lambda \rightarrow \infty $, we observe $p_{(\bullet \circ \bullet )\rightarrow (\bullet \bullet \bullet )}\rightarrow 1$, $p_{(\circ \circ \bullet )\rightarrow (\circ \bullet \bullet )}=p_{(\bullet \circ \circ )\rightarrow (\bullet \bullet \circ )}\rightarrow 1/2$, and $p_{(\bullet )\rightarrow (\circ )}\rightarrow 0$.
The system tends to evolve toward a state where all sites are active. On the other hand, when $\lambda \rightarrow 0$, $p_{(\bullet \circ \bullet
)\rightarrow (\bullet \bullet \bullet )}=p_{(\circ \circ \bullet
)\rightarrow (\circ \bullet \bullet )}=p_{(\bullet \circ \circ )\rightarrow
(\bullet \bullet \circ )}\rightarrow 0$ and $p_{(\bullet )\rightarrow (\circ
)}\rightarrow 1$, causing all sites to become inactive.

In the Ising model, for instance, at $T\rightarrow \infty $, $%
p(s_{i}=+1)=p(s_{i}=-1)=\frac{1}{2}$, and the time evolution of the
magnetization, $m(t)=\frac{1}{N}\sum\nolimits_{i=1}^{N}s_{i}(t)$, follows a
Gaussian distribution that is uncorrelated with any other evolution.
Therefore, in this case, we expect the system to recover the MP law, as
given by Eq. (\ref{Eq:MP}). However, in the CP model, a gap dynamics (different from Ising systems) must still be maintained to ensure that the minimum average eigenvalue reaches a minimum at $\lambda =\lambda _{c}$.

\section{Results}

\label{Sec:Results}

Now we present our results for the estimates of the critical value as a
function of mobility using both methods previously described: Method 1:
Optimization of the power laws, and Method 2: Spectral analysis of
correlation matrices.

\subsection{Method 1}

We begin by performing MC simulations to determine the
critical parameter for different diffusion settings. The range of $\lambda
_{\min }$ spans from $1.5$ to $3.5$, with a resolution of $\Delta \lambda
=0.01$. We simulate the contact process with mobility for systems of size $%
L=1024$, running a total of $N_{steps}=500$ MC steps for each evolution. The
first $N_{\min }=30$ MC steps are discarded, and we then average the results
of $\left\langle \rho (t|\lambda ,\gamma )\right\rangle $ over $N_{run}=2000$
different time evolutions for each mobility value $\gamma $. Just as a test of our approach and fast comparison with results presented in literature, when considering $\gamma =0$, we obtain $\delta =\frac{\beta }{\nu _{\parallel }}\approx 0.159$ for a single seed, which is very close to the known estimates to the one-dimensional directed percolation $\delta =0.159464(6)$ \cite{Hinrichsen2000}.

Next, we optimize the process by calculating the coefficient of
determination $r$ obtained from the linear fit of $\left\langle \rho
(t|\lambda ,\gamma )\right\rangle \times t$ in log-log scale for each of the
200 $\lambda $--values tested, across 11 different values of $\gamma $,
ranging from $\gamma =0.0$ to $\gamma =1.0$ ($\Delta \lambda =0.1$). The
results for $r$ as a function of $\lambda $ are shown in Figure \ref%
{fig:r_x_gamma}.

\begin{figure}[tbh]
\begin{center}
\includegraphics[width=1.0\columnwidth]{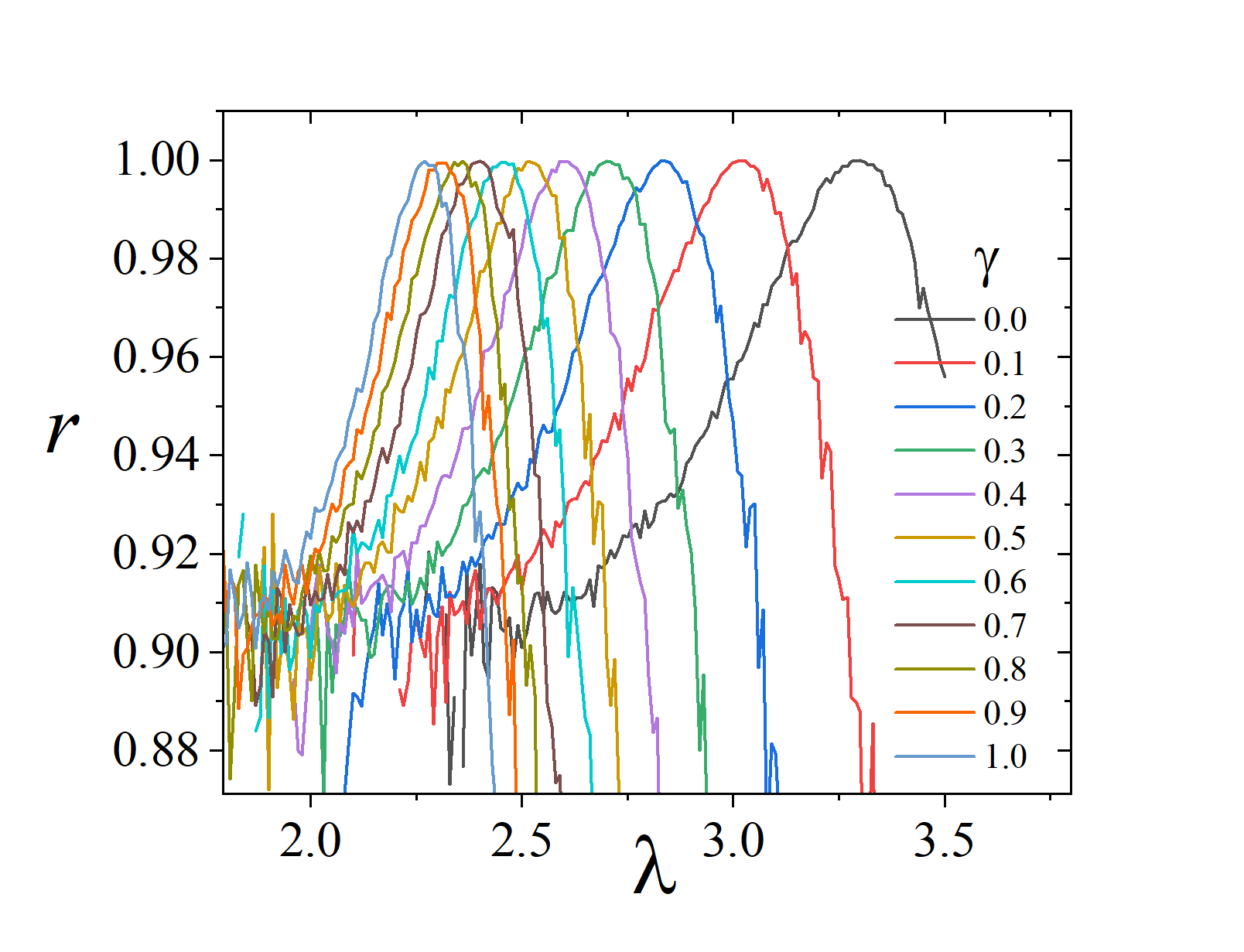}
\end{center}
\caption{Coefficient of determination, $r$, as a function of $\protect%
\lambda $ for $1.5\leq \protect\lambda \leq 3.5$ with $\Delta \protect%
\lambda =10^{-2}$. The different curves correspond to values of $\protect%
\gamma $ ranging from $0$ to $1.0$, with increments of $0.1$ (i.e., $\protect%
\gamma =0,0.1,0.2,...,0.9,1.0$). }
\label{fig:r_x_gamma}
\end{figure}

As observed, $\lambda _{c}$ decays as a function of $\gamma $, in
(qualitative) agreement with the prediction of the mean-field approximation,
although not exhibiting the same simple behavior (i.e., $\lambda _{c}=1-D$)
found there. It is important to note that this is not directly comparable,
as the dependence of $\gamma $ on $D$ is not known. Regarding the fits of $%
\lambda _{c}$ as a function of $\gamma $, relevant fits will be discussed in
Subsection \ref{Subsec:Results_Comparison}. First, we will demonstrate that
Method 2 corroborates the results obtained from Method 1, showing that both
methods can be used to localize the critical parameters of the contact
process (CP) model with mobility.

\subsection{Method 2}

\label{Subsec:Results_Method2}

In the second method, we construct correlation matrices with $%
N_{sample}=100$ different time evolutions. Unlike the approach used for the Ising model as well as others up-down symmetry models, we set $\rho _{0}=1$ here. It is
important to note that in this case, there is no order-disorder transition,
and the minimal density for evolution is $\rho _{0}=1/L$, which has a
different effect compared to $\left\langle m_{0}\right\rangle =0$ in the
up-down symmetry models.

\begin{figure*}[tbh]
\begin{center}
\includegraphics[width=1.2\columnwidth]{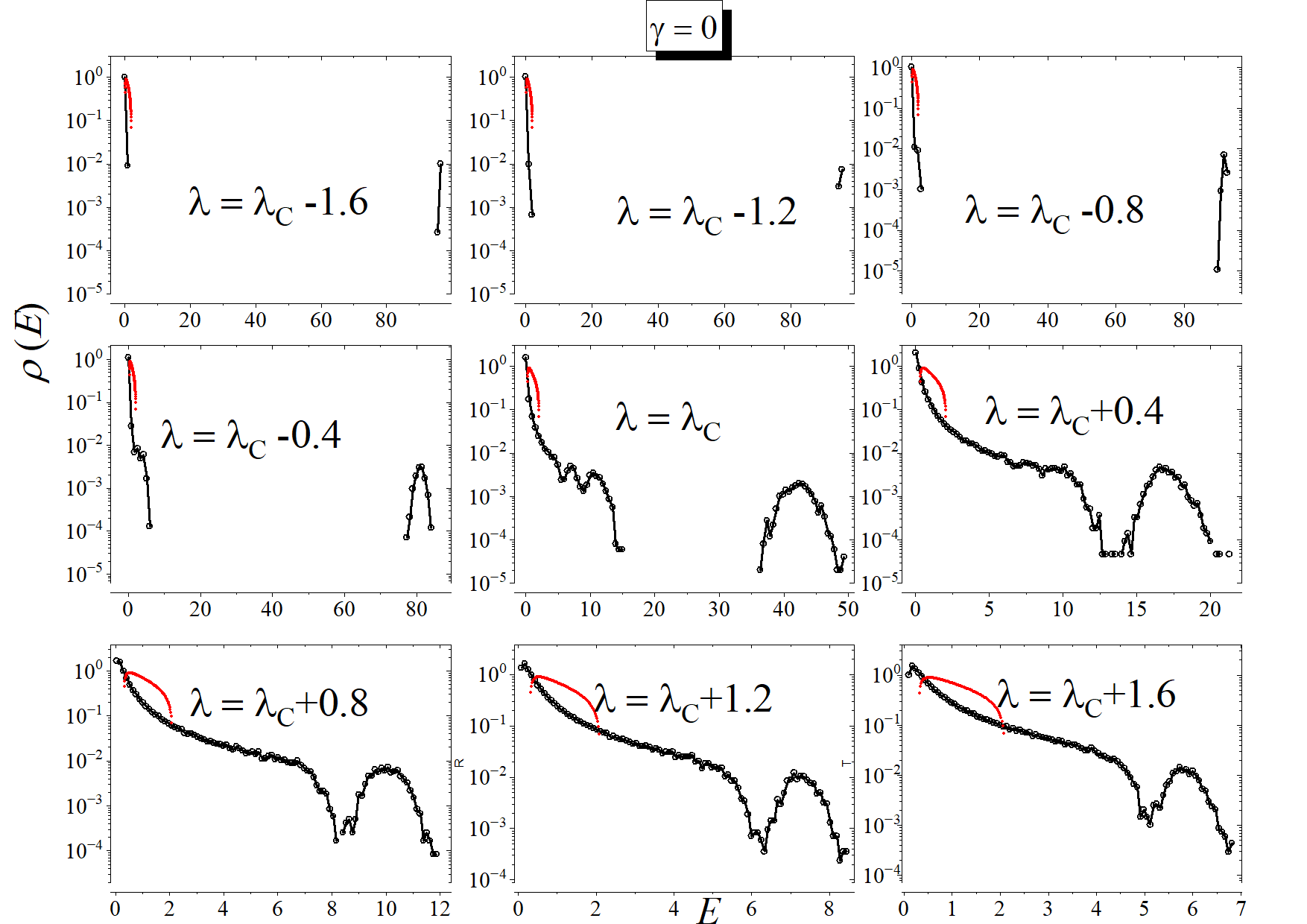}\\[0pt%
]
\includegraphics[width=1.2\columnwidth]{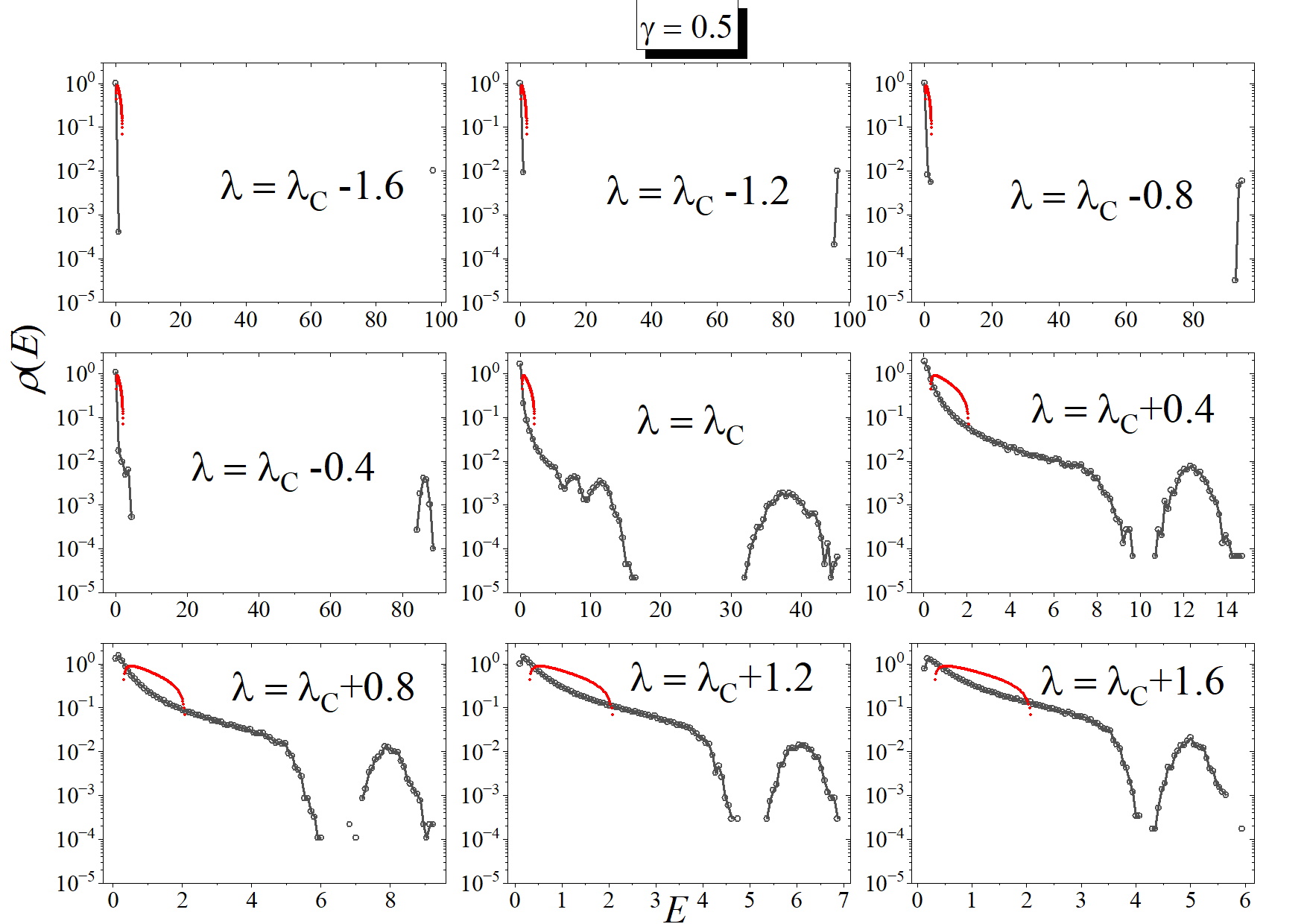}\\[%
0pt]
\includegraphics[width=1.2\columnwidth]{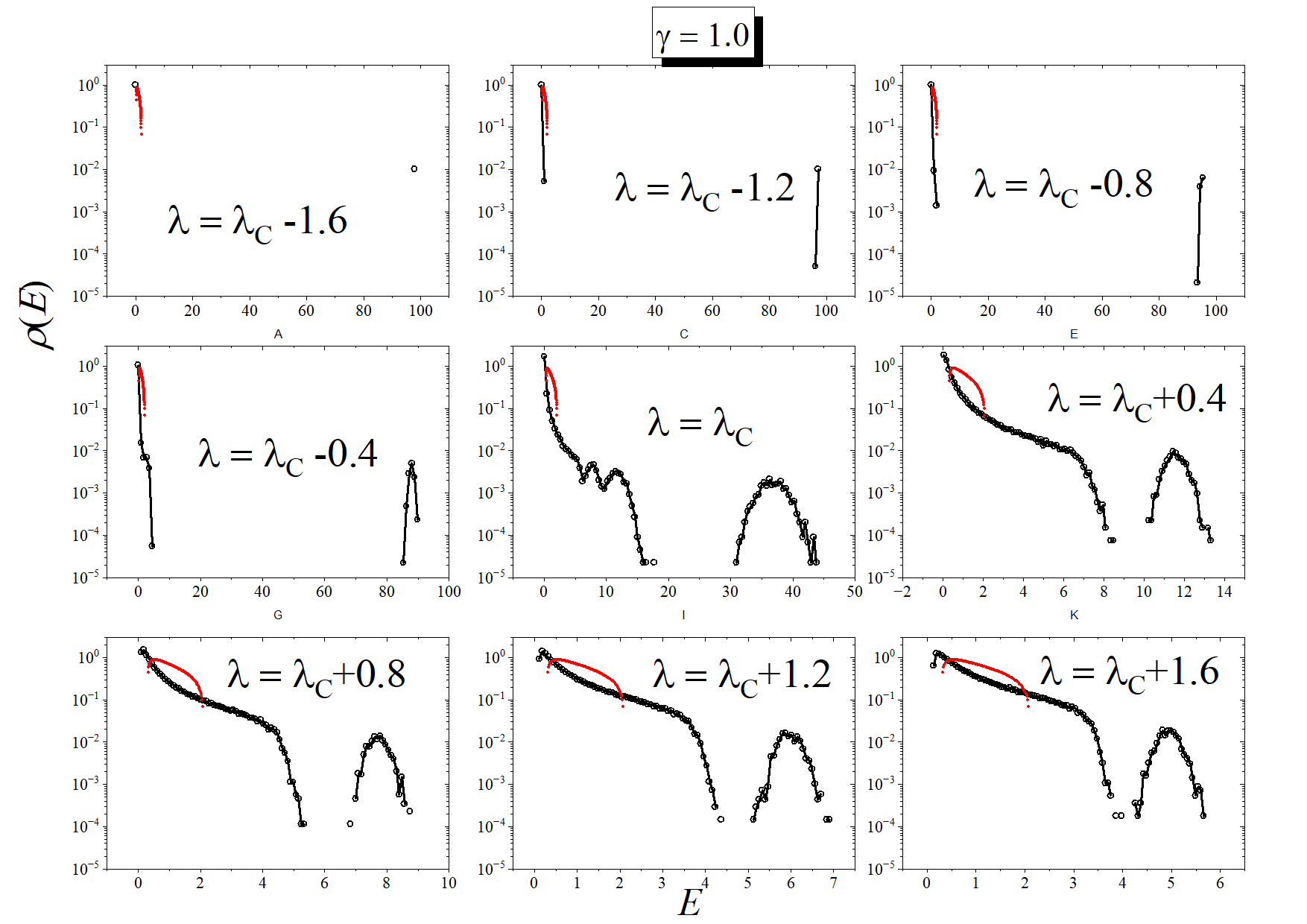}
\end{center}
\caption{The density of states for $\protect\gamma =0$, $\protect\gamma =0.5$%
, and $\protect\gamma =1.0$ is shown for different values of $\protect%
\lambda $. The red curves represent the Marchenko-Pastur (MP) law for $%
N_{steps}=500$ and $N_{sample}=100$ for comparison, highlighting that,
particularly in the contact process, the MP law does not play a significant role, differing from what is observed for kinect Ising systems}.
\label{Fig:density_of_states}
\end{figure*}

In addition, we built histograms from $N_{run}=1000$ different random matrices by considering $L=2^{7}$ sites (fewer than in the previous method, which is consistent with the results obtained in Ref. \cite{RMT2023} -- see Fig. 3 of that this reference) and $N_{steps}=500$, yielding a total of $10^{5}$ eigenvalues. Simulations were
performed for different values of $\gamma $. It is interesting to observe
how the density of states, $\sigma (E)$, behaves for various values of $
\lambda $\footnote{It is important to construct these histograms correctly, rather than calculating the averages directly, for the method to work properly.}. Unlike observed in up-down symmetry systems, where high temperatures lead to a migration of the density of states toward the MP law, in the contact process the result shows that migration does not occur, as can be seen in Fig. \ref{Fig:density_of_states} for different values of $\gamma$.

As previously noted, when $\lambda \rightarrow \infty $ or $\lambda
\rightarrow 0$, the system exhibits two distinct absorbing states due to its
symmetry. In one case, all particles are active, while in the other, all
particles are inactive. In both scenarios, the system necessarily deviates
from the MP law. The red curves in the plots demonstrate that
the density of states significantly differs from the MP law in these
situations, a characteristic that distinguishes this system from Ising-like
models. 

Although there is a gap structure in the eigenvalues that resembles that of Ising-like models, we cannot establish a direct link between these gaps and the phase transition simply by observing the eigenvalue bulk and gap dynamics. It is important to note that even in Ising-like systems, the dynamics of gap closing alone does not fully explain phase transitions based on these matrices. However, fluctuations in the histograms (density of states) of both Ising-like models \cite{RMT2023} and the current contact process seem to reliably predict the existence of critical points, regardless of our understanding of the bulk and gap dynamics. This observation warrants a more in-depth theoretical exploration, though this is beyond the scope of the current study.

\begin{figure*}[tbh]
\begin{center}
\includegraphics[width=1.0\columnwidth]{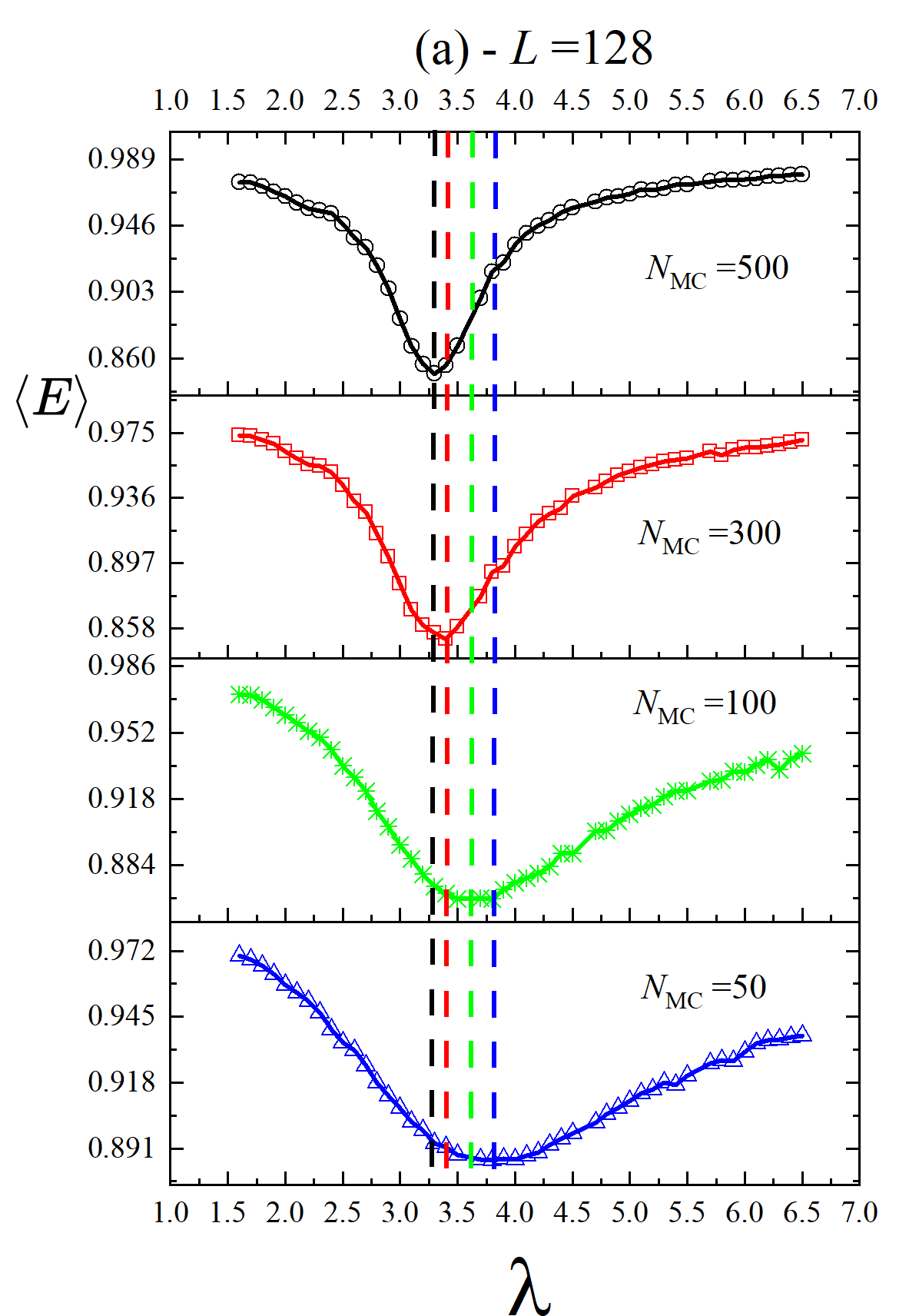}%
\includegraphics[width=1.0\columnwidth]{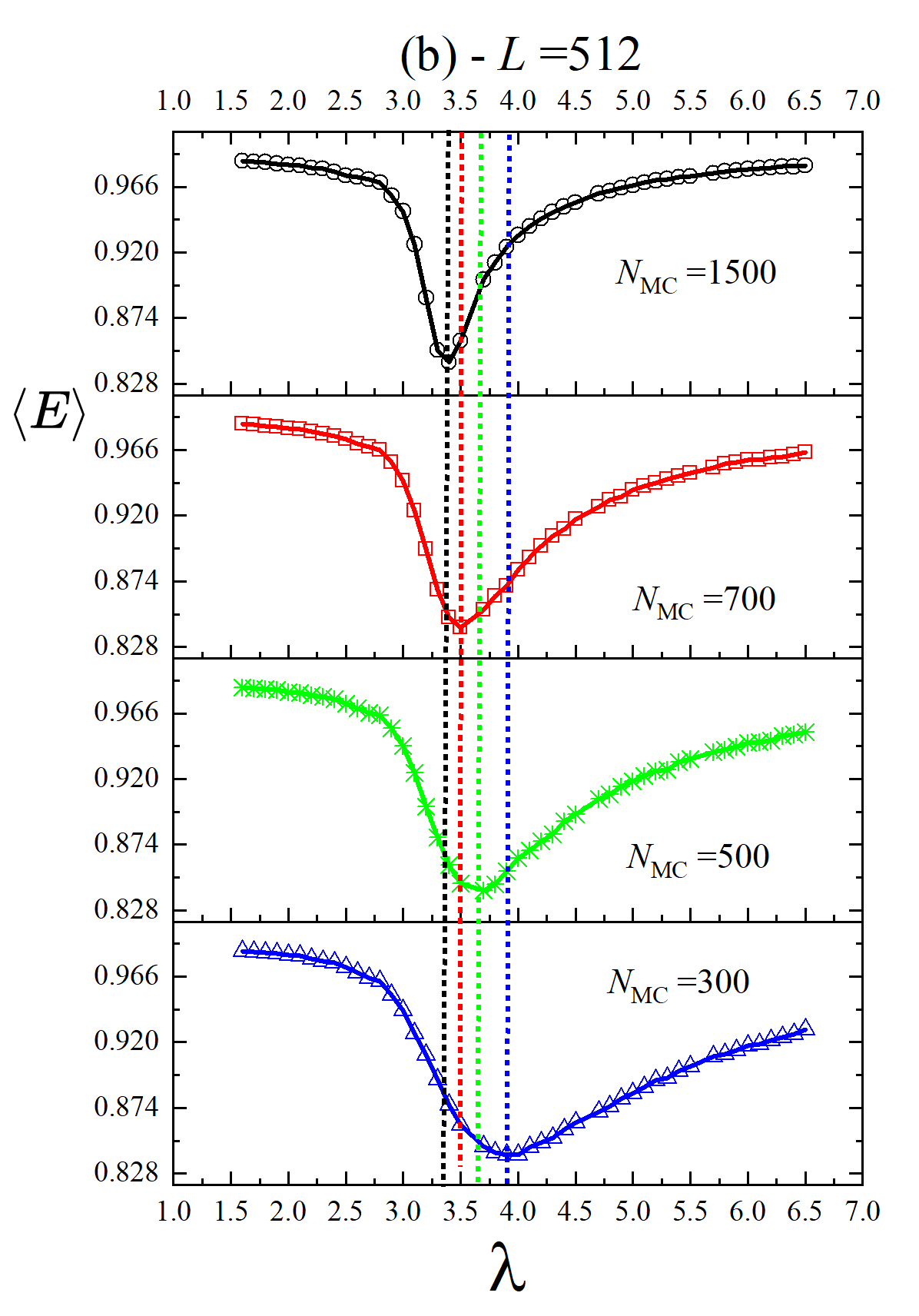}
\end{center}
\caption{Average eigenvalue $\left\langle E\right\rangle $ as a function
of $\protect\lambda $ is shown for two lattice sizes: (a) $L=128$ and (b) $%
L=512$. For each lattice size, we vary the number of Monte Carlo steps, $%
N_{steps}$. For $L=128$ and $N_{steps}=500$, we obtain a result of $\protect%
\lambda \approx 3.29$, which is already very close to the critical point for 
$\protect\gamma =0$. In contrast, for $L=512$, a significantly higher number
of steps, $N_{steps}$ $=1500$, is required to achieve a similar level of
accuracy, indicating a much higher computational cost.}
\label{Fig:LversusNsteps}
\end{figure*}

For example, refer to Fig. \ref{Fig:LversusNsteps}, where we used the
density of states to calculate the average eigenvalue $\left\langle
E\right\rangle $ as a function of $\lambda$ for two lattice sizes: (a) $L=128$
and (b) $L=512$, with $\gamma =0$, without loss of generality. We observe
that the method works well for small systems, as we used only $L=2^{7}$
sites, with results depending on the adjusted $N_{steps}$. Larger systems,
however, require significantly more computational effort, as indicated by
Fig. \ref{Fig:LversusNsteps} (b). Based on Eq. (\ref{Eq:Average}), we analyze
the results for $L=128$ and $L=512$ sites. Drawing from the hypothesis
presented in previous works \cite{RMT2023,RMT2023-4}, where a minimum in the
average eigenvalue suggests the existence of a critical point, we explore
how this method performs for these two system sizes.

For each lattice, we varied $N_steps$. Observations show that for $L=128$
with $N_{steps}=500$ Monte Carlo (MC) steps, we obtained $\lambda \approx
3.29$, a value very close to the critical point expected for $\gamma =0$.
However, for $L=512$, a significantly higher computational cost was
required, with $N_{steps}=1500$ MC steps. This suggests that using $L=512$
is unnecessary, as the correlation matrix method works efficiently for $L=128
$. The problem is well-calibrated for this system size using $N_{steps}=500$
MC steps.

\begin{figure*}[tbh]
\begin{center}
\includegraphics[width=1.0%
\columnwidth]{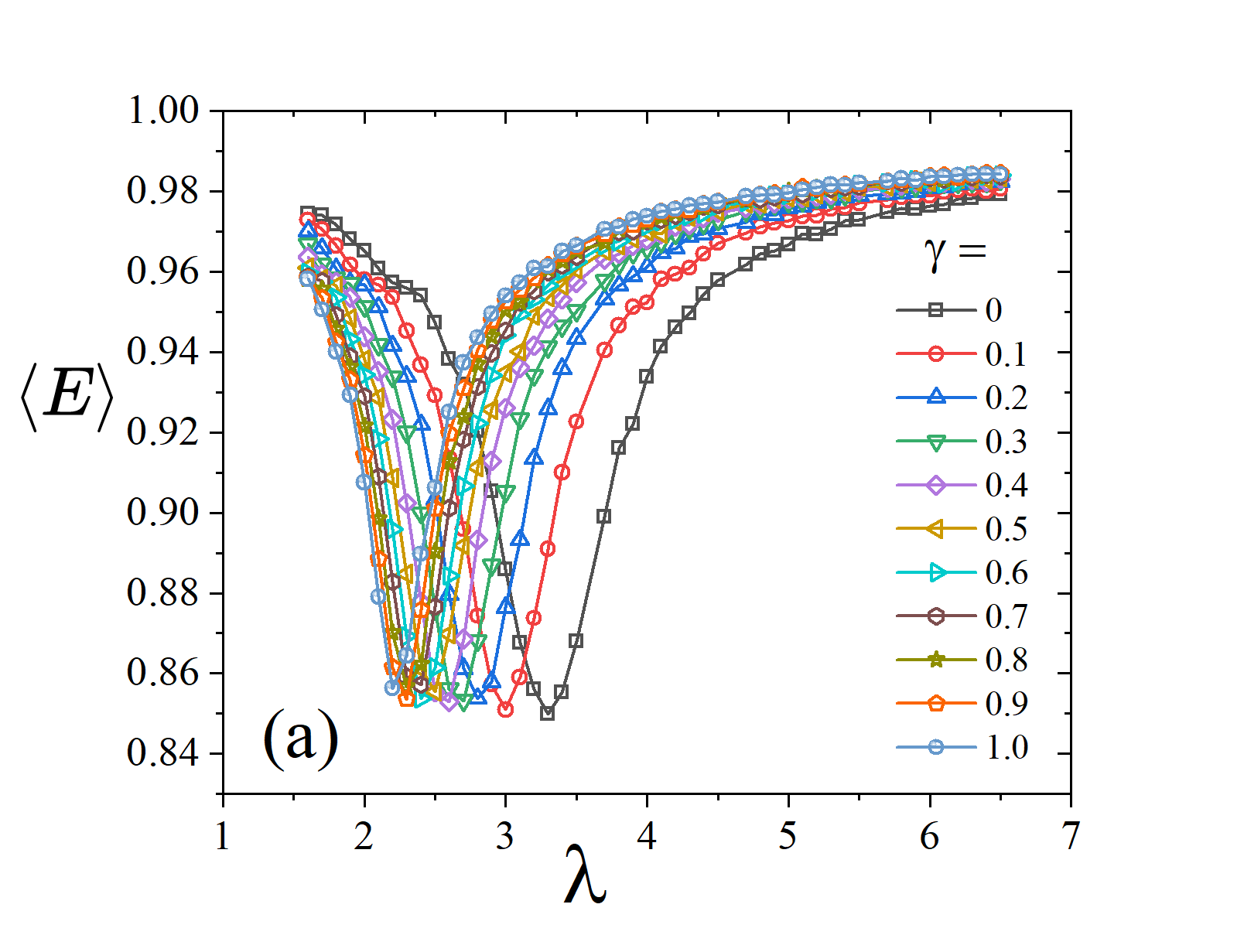}%
\includegraphics[width=1.0\columnwidth]{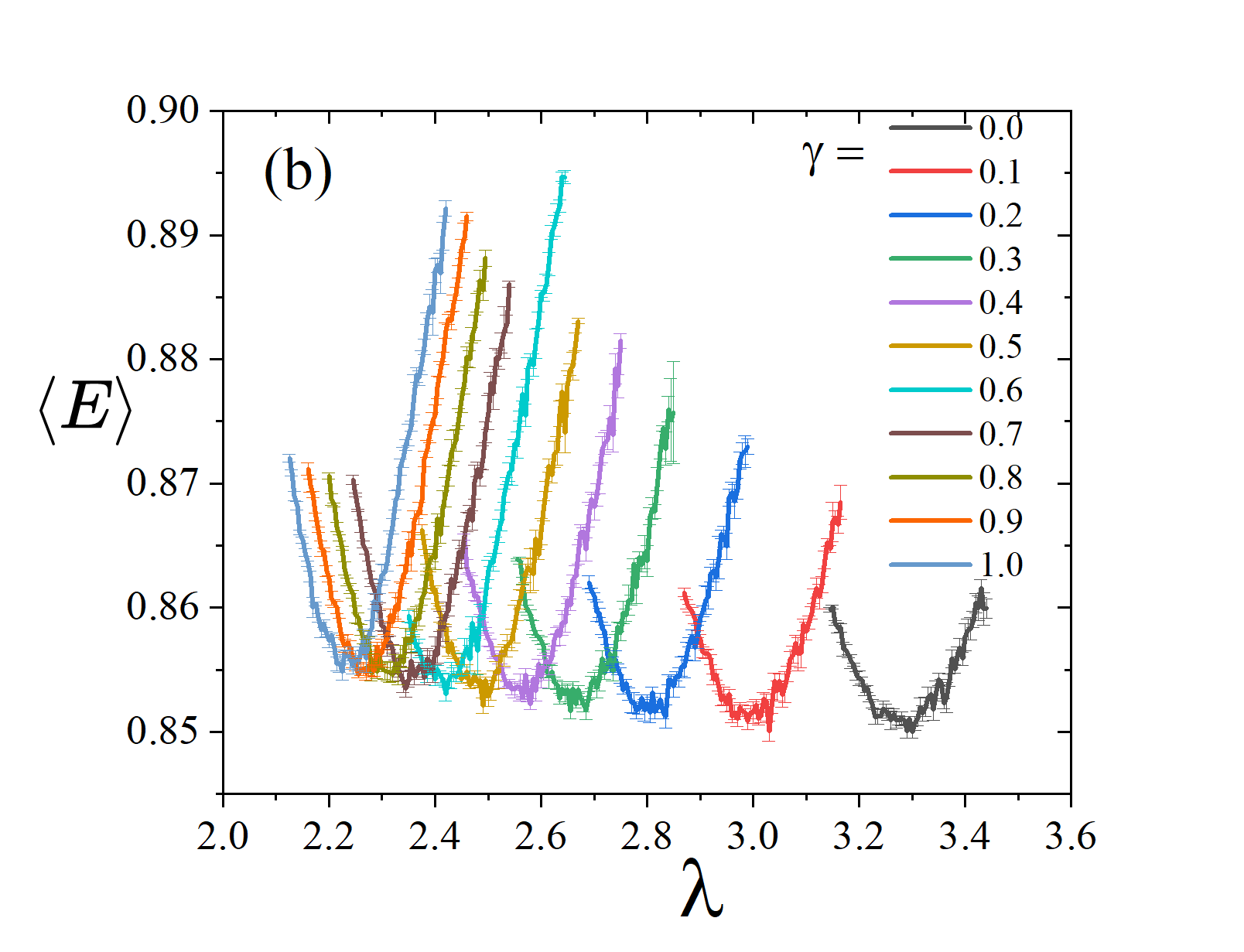}
\end{center}
\caption{(a) Average eigenvalue $\left\langle E\right\rangle $ as a
function of $\protect\lambda $ for various values of $\protect\gamma $ during the coarse-grained stage with $\Delta \protect\lambda =0.1$. (b) Average eigenvalue $\left\langle E\right\rangle $ as a function of $\protect\lambda $ for various $\protect\gamma $ values in the fine-scale regime with $\Delta \protect\lambda =0.005$.}
\label{Fig:average_eigenvalue_coarse_and_fine}
\end{figure*}

Thus, for $L=128$, we calculated $\left\langle E\right\rangle $ as a
function of $\lambda $ for various $\gamma $ values. Initially, we used $%
\Delta \lambda =0.1$, referred to as the coarse-grained stage. The minimal
value of $\left\langle E\right\rangle $ indicated that the critical
temperature varies as $\gamma$ changes (see Fig. \ref{Fig:average_eigenvalue_coarse_and_fine} (a)).

After an initial rough localization, we then proceed with a finer
refinement using $\Delta \lambda =0.005$ (fine-scale stage). The behavior of $\left\langle E\right\rangle $ as a function of $\lambda $ is presented in Fig. \ref{Fig:average_eigenvalue_coarse_and_fine} (b). Despite averaging across five different seeds, the process remains notably noisy. However, we successfully determined the minimum point, $\lambda _{c}$, by averaging results from the different seeds. These findings will now be compared to those obtained using the power-law optimization method (Method 1).

\subsection{Comparison between the methods}

\label{Subsec:Results_Comparison}

\begin{figure}[tbh]
\begin{center}
\includegraphics[width=1.0\columnwidth]{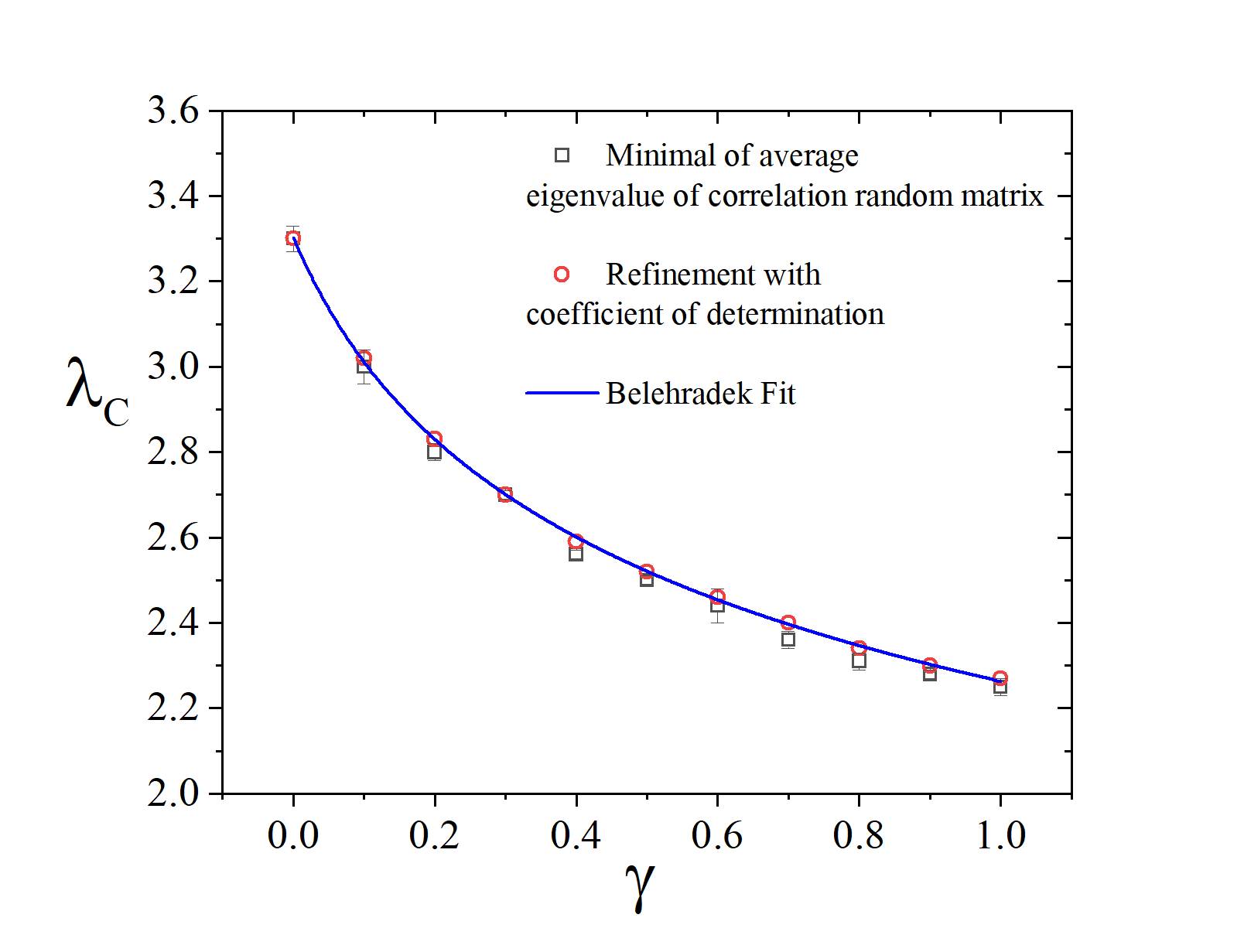}
\end{center}
\caption{Comparison between Method 1 (optimization of power laws) and Method
2 (spectral analysis of correlation matrices). The results are in agreement
and also appear to be well-fitted to the Belehradek fit.}
\label{Fig:comparison}
\end{figure}

In Fig. \ref{Fig:comparison}, we show $\lambda _{c}$ as a function of $%
\gamma $ using the two methods described in this manuscript. The results
appear to be in good agreement. The square markers represent the method
where we calculated the minimum of the average eigenvalue of correlation
matrices constructed from the time evolution of the density of active sites
(Method II), while the circular markers correspond to the method where we
observed the temporal power-law averages across different time evolutions.

We observe a non-trivial decay of $\lambda _{c}$ with $\gamma $, which
differs from the behavior seen in the mean-field regime, where $\lambda
_{c}=1-D$. After empirically testing several candidate functions, we found
that the Belehradek function, widely known in the literature, provides a
good fit:

\begin{equation*}
\lambda _{c}=a(\gamma -b)^{c}
\end{equation*}
where we determined $a\approx 2.33$, $b\approx -0.16$, and $c\approx -0.19$,
yielding an excellent fit quality with a coefficient of determination (COD)
of approximately $0.999667$. 

The Belehradek equation is traditionally associated with temperature effects
in biological processes (e.g., see \cite{Ross1993}). In this context, we
adopt it in an entirely empirical manner. We could potentially interpret the
parameter $\gamma \ $as a form of mobility probability associated to the
temperature of the system, suggesting that it influences the critical
behavior of living matter, although this interpretation is totally
qualitative.

The key point here is that both methods are in agreement, providing a strong
fit that describes how the critical parameters of the contact process change
with mobility, which follows the Belehradek function. This suggests that the
results qualitatively support our simple mean-field model, where $\lambda
_{c}$ decays linearly with $D$, a parameter that is related to the mobility
of the system $\gamma $.

\section{Conclusions}

\label{Sec:Conclusions}

We studied the contact process using two nonequilibrium methods:
optimization of temporal power laws and spectral analysis of correlation
matrices, both based on $N_{sample}$ evolutions of the density of active
particles. Specifically, we examine how the critical rate $\lambda _{c}$
varies with the mobility of particles $\gamma $. Our results demonstrate
that $\lambda _{c}$ decays with $\gamma $ following a Belehradek function, with both methods presenting good agreement.

These findings qualitatively support the mean-field regime, where $\lambda_{c}$ is expected to decay with mobility, which in this case is represented by the parameter $D$.

\section*{Acknowledgments}
R. da Silva would like to express gratitude to M. J. de Oliveira and R. Dickman for 
their valuable insights regarding the contact process. Additionally, he acknowledges the 
financial support provided by CNPq under grant number 304575/2022-4

\end{document}